\documentclass[preprint,preprintnumbers,amsmath,amssymb, prl]{revtex4}
\usepackage[pdftex]{graphicx}
\usepackage{bm}
\usepackage{float}
\usepackage{color}
\usepackage{hyperref}
\usepackage{natbib}
\usepackage[T1]{fontenc}
\usepackage{wasysym}
\definecolor{heartgold}{rgb}{0.5, 0.5, 0.0}
\usepackage{txfonts}
\DeclareFontEncoding{LS1}{}{}
\DeclareFontSubstitution{LS1}{stix}{m}{n}
\DeclareSymbolFont{symbols4}{LS1}{stixbb}{m}{it}
\DeclareMathSymbol{\varhexagonblack}{\mathord}{symbols4}{"DD}
\DeclareMathSymbol{\hexagonblack}   {\mathord}{symbols4}{"DE}
\definecolor{navyblue}{rgb}{0.0, 0.0, 0.5}
\definecolor{darkpastelpurple}{rgb}{0.59, 0.44, 0.84}
\definecolor{purple(html/css)}{rgb}{0.5, 0.0, 0.5}
\definecolor{violet}{rgb}{0.56, 0.0, 1.0}

\begin{document}
	
	\title{\color{blue}Influence of dispersion medium structure on the physicochemical properties of aging colloidal suspensions investigated using the synthetic clay Laponite\textsuperscript \textregistered
	}
	
	\author{Chandeshwar Misra}
	\affiliation{Soft Condensed Matter Group, Raman Research Institute, C. V. Raman Avenue, Sadashivanagar, Bangalore 560 080, INDIA}
	\author{Venketesh T Ranganathan}
	\affiliation{Soft Condensed Matter Group, Raman Research Institute, C. V. Raman Avenue, Sadashivanagar, Bangalore 560 080, INDIA}
	\author{Ranjini Bandyopadhyay}
	\email{ranjini@rri.res.in}
	\affiliation{Soft Condensed Matter Group, Raman Research Institute, C. V. Raman Avenue, Sadashivanagar, Bangalore 560 080, INDIA}
	
	\date{\today}
	\definecolor{carmine}{rgb}{0.59, 0.0, 0.09}
	\begin{abstract}
		\paragraph{}
		{\bf Hypothesis:} Aging in colloidal suspensions manifests as a reduction in kinetic freedom of the colloids. In aqueous suspensions of charged colloids, the role of inter-particle electrostatics interactions on the aging dynamics is well debated. Despite water being the dispersion medium, the influence of water structure on the physicochemical properties
		of aging colloids has never been considered before. Laponite, a model hectorite clay, could be used to evaluate the relative contributions of medium structure and electrostatics in determining the physicochemical properties of aging colloidal suspensions.
		\paragraph{}
		{\bf Experiments:} The structure of the dispersion medium is modified either by incorporating uncharged/charged kosmotropic (structure-inducing) or chaotropic (structure-disrupting) molecules or by changing suspension temperature. A new protocol, wherein the medium is heated before adding clay particles, is also introduced to evaluate the effects of hydrogen bond disruptions on suspension aging. Dynamic light scattering, rheological measurements and particle-scale imaging are employed to evaluate the physicochemical properties of the suspensions.
		
		\paragraph{}
		{\bf Findings:} A strong influence of medium structure is evident when inter-particle electrostatic interactions are weak. Enhancement and disruption of hydrogen bonds in the medium are, respectively, strongly correlated with acceleration and delay of suspension aging dynamics. The physicochemical properties of charged clay colloidal suspensions are therefore controlled by altering hydrogen bonding in the dispersion medium.
		
		\paragraph{}	
		{\bf Key words :} Nano-clay colloids; Dispersion medium structure; Suspension structure and dynamics; Inter-particle electrostatic interactions; Rheology, Dynamic light scattering; Cryogenic scanning electron microscopy
		
		\paragraph{}	
		{\bf Abbreviations :} EDL, Electrical Debye/double layer; DMF, N,N-Dimethylformamide; DLS, Dynamic light scattering; LVE, Linear viscoelastic; cryo-SEM, cryogenic scanning electron microscopy; Glu, Glucose; OC, Overlapping coin; HOC, House of cards. 	
	\end{abstract}
	
	\maketitle     
	%
	%
	\section{1. Introduction}
	Aging in colloidal suspensions imposes dynamical constraints on the particles leading to their kinetic arrest \cite{B_Ruzicka_2011,D_Bonn_1999,D_Saha_2014}. In charged colloidal suspensions, the aging dynamics and the subsequent changes in the physicochemical properties of the system are attributed to the growth of inter-particle electrostatic interactions. Charged clay colloids serve as excellent model candidates to investigate the aging dynamics and the physicochemical properties of the sample \cite{A_Mourchid_1995,D_Saha_2015}. The structure \cite{B_Ruzicka_2011,S_Jabbari-Farouji_2008}, dynamics \cite{R_Bandyopadhay_2004,D_Saha_2015_1,K_Suman_2018,B_Zheng_2020} and flow properties \cite{A_Mourchid_1998,R_Bandyopadhay_2010,G. Mallikarjunachari_2018,Hossein_A_Baghdadi,A_S_Negi_2010} of suspensions of smectite clay minerals have been studied extensively in the literature. Sodium-montmorillonite, also referred to as sodium bentonite, and kaolinite are two examples of clay minerals belonging to the smectite group that are available in abundance in nature \cite{F. Bergaya_2013}. The phase and bulk behaviors of smectite clay minerals can be easily tuned by tuning inter-particle interactions. This has led to the widespread applications of these clay minerals as rheological modifiers and stabilizers in paints, wellbore drilling fluids, cosmetics, pharmaceuticals, agrochemicals, paper fillers, coating pigments and nanocomposites \cite{B. Additives_2014,F_Uddin_2008,H_H_Murray_2000}.
	\paragraph{}
	Laponite, belonging to the class of hectorite clays, exists as stacks of tactoids. The structure of individual Laponite particles and the phase behavior of their suspensions are described elsewhere \cite{J_Madejova_1998, S_L_Tawari_2001, M. Delhorme_2012,M. Delhorme_2014}. When Laponite clay is dispersed in an aqueous medium, water molecules diffuse into the intra-gallery spaces of the clay platelets and hydrate the sodium counterions, thereby causing the tactoids to swell and exfoliate \cite{Y_M_Joshi_2007}.
	Due to the gradual buildup of osmotic pressure gradients in the sample, the sodium ions diffuse into the bulk, creating excess negative charges on the Laponite surface \cite{S_Ali_2013, D_Saha_2015}. This is accompanied by the formation of an electrical double layer that evolves with time due to the time-dependent development of inter-particle interactions. The positively charged EDLs repel strongly and this leads to kinetic arrest of the Laponite platelets at long waiting times, $t_{w}$. The aging dynamics and self-assembled structures formed in Laponite suspensions are, therefore, attributed to the growth of time dependent inter-particle electrostatic interactions \cite{B_Ruzicka_2011,S_Ali_2013, B_Zheng_2020,Hossein_A_Baghdadi,A_S_Negi_2010,A_Mourchid_1998,B_Ruzicka_2004,T_Nicolai_2001,P_Mongondry_2005,B_Abou_2001}. 
	The physicochemical properties of Laponite suspensions can be effectively tuned  by controlling the inter-particle interactions using external additives, applied electric field and by varying the suspension temperature \cite{V_Ranganathan_2017,D_Saha_2015,P_Gadige_2018}.
	\paragraph{}
	Soil, in which clay is an important ingredient, and water are vital factors in sustaining life on earth. On the one hand, water and salinity in soil influence the production of agricultural crops \cite{Soil_veg, soil_water_content}, while on the other hand, they control large scale geophysical phenomena such as river delta formation, landslides and earthquake-driven liquefaction \cite{A_Thill_2001,A_Khaldoun_2005,Marika_Santagata_2014}. Na-montmorillonite nanoclay is often used as a laboratory model clay to model these geophysical phenomena \cite{Samim_2015}. Water is a structured medium in which pairs of water molecules engage in hydrogen bonding and arrange in tetrahedral configurations, the most energetically favored structure of liquid water \cite{R_Ludwig_2001, J_D_Smith_2005}. Despite extensive investigations on the effects of water content and salinity on the structure and dynamics of soil, the influence of the structure of water, achieved through controlled variations in hydrogen bond populations, on the physicochemical properties of clay suspensions has never been reported to the best of our knowledge. The contributions of inter-particle electrostatic interactions, in contrast, have been extensively debated in the literature.
	\paragraph{}
	In the present study, we focus our attention on the relative contributions of dispersion medium structure and electrostatic interactions on the physicochemical properties of aging suspensions of aqueous Laponite clay colloids. To enhance the hydrogen bond population in liquid water, we employ glucose, a kosmotropic molecule (a molecule inducing structure formation in the aqueous medium) which form dense layers of structured water around the molecule \cite{J_W_Brady_1989, M_Paolantoni_2007}. In another experiment, we disrupt the water structure by adding N,N-Dimethylformamide (DMF), a chaotropic molecule (a molecule that disrupts the structure of the aqueous medium) to the dispersion medium. DMF disrupts hydrogen bonding in liquid water by forming hydrogen bonds with the oxygen sites of DMF \cite{Y_Lei_2003,J_M_M_Cordeiro,P_M_Geethu_2018}. We suspend Laponite in aqueous solutions of glucose and DMF prepared at desired concentration in order to uncover the role of structure of the dispersion medium on the suspension microstructure, dynamics and bulk flow properties. We hypothesize that since glucose and DMF are non-dissociating, uncharged molecules, these experiments should reveal the influence of structure of the dispersion medium on the physicochemical properties of clay suspensions. It is reported that the osmotic pressure of a solution strongly depends on the formation of hydration shells of water around the solute and anion-cation interactions in the ionic solutions \cite{J_Cannon_2012,Y_Luo_2010}. Since dense layers of water molecules form around glucose molecules, osmotic pressure of the solution is enhanced \cite{M_A_Darwish_2014,V_Gekas_1998}. In contrast to glucose, DMF molecules disrupt the water structures and water molecules do not form hydration shells around DMF molecules. This decreases the osmotic pressure of an aqueous solution \cite{P_Wiggins_2008} in the presence of DMF. It is well-established in the literature that osmotic pressure gradients are the driving force behind suspension aging \cite{Y_M_Joshi_2007,A_Mourchid_1995}. By performing systematic experiments, we report here that the presence of a kosmotropic molecule in solution accelerates the aging dynamics in clay suspensions, while these dynamics are retarded when a chaotrope is added. 
	\paragraph{}
	While the addition of uncharged non-dissociating molecules reveals the role of dispersion structure, charged dissociating molecules that can alter the local water structure can be employed to evaluate the competition between dispersion structure and electrostatic interactions on the suspension properties. We implement this experimentally by adding two charged dissociating molecules, NaCl, a kosmotrope \cite{P_M_Geethu_2018,P_Ben_Ishai_2013,X_Ma_2006}, and KCl, a chaotrope, to study their influence of suspension properties. The action of NaCl and KCl on the local structure of water is described elsewhere \cite{P_Ben_Ishai_2013, X_Ma_2006, M_A_Darwish_2014}. Regardless of whether the dissociating molecules have a kosmotropic or chaotropic effect on the local structure of the dispersion medium, we observe only accelerated suspension aging. We attribute this to the dominance of strong inter-particle electrostatic interactions \cite{D_Saha_2015,K_Suman_2018}, induced by dissociating molecules, over changes in dispersion medium structure.
	\paragraph{}
	Finally, we use temperature as an additional control parameter to disrupt hydrogen bonds in liquid water. We device a new experimental protocol, where we preheat the dispersion medium to ensure the de-structuring of hydrogen bonds before adding Laponite clay particles (pre-heating). In another sets of experiments, we heat the suspensions after the addition and mixing of clay particles has been completed (post-heating). Besides disrupting hydrogen bonds, an increase in temperature of aqueous Laponite suspensions is expected to shrink the electrical double layer due to an increase in the number of dissociated Na$^+$ ions in the medium \cite{D_Saha_2015}. We report enhancements in the dissociation of Na$^+$ and the suspension aging dynamics when the dispersion medium temperature is raised before the addition of Laponite particles (pre-heated experiments), when compared to experiments in which the suspension is heated after the addition of Laponite (post-heated experiments). These experiments confirm that electrostatic interactions dominate over any changes in the local structure of the dispersion medium in determining the physicochemical properties of suspensions in the presence of charged dissociating molecules and when temperature is raised.
	\paragraph{}
	The microscale data uncovered from our dynamic light scattering studies are supported by rheological measurements. Particle-scale images are  acquired using cryogenic electron microscopy and the morphologies of the observed suspension structures are quantified through measurements of pore areas and network branch thicknesses. Regardless of the predominance of electrostatics or dispersion medium structure on the properties of the suspensions, the microscopic aging dynamics are self-similar in the presence of all the additives and at all suspension temperatures explored here. We further report that viscoelastic moduli under large strains are sensitive to the presence of additives and temperature histories. The novelty of this study lies in the identification of the dispersion medium structure as a key parameter in determining the physicochemical properties of aging colloidal suspensions. Our results could be expanded to any aqueous aging system in which osmotic pressure gradients and electrostatics control the physicochemical properties.

	\section{2. EXPERIMENTAL SECTION}
	\subsection{2.1. Sample preparation}
	Laponite XLG\textsuperscript \textregistered powder (purchased from BYK Additives Inc., molecular weight- 2286.9 g/mol) was dried in an oven at $120$$^{\circ}$C for 18 h to remove moisture. The dried powder was weighed and added to Milli-Q water (Millipore Corp., resistivity 18.2 M$\Omega$-cm) which was continually agitated using a magnetic stirrer for 45 min. The additives glucose (Sigma-Aldrich), DMF (SDFCL  Fine Chem Pvt.Ltd), NaCl (LABORT Fine Chem Pvt.Ltd.), and KCl (Sigma-Aldrich) were measured and added to de-ionized water prior to the addition of Laponite powder. All the additives were used as received without any further purification. Experiments at different suspensions temperatures were performed following two different protocols. In the first protocol, the temperature of the water was set before the addition of Laponite powder and was maintained throughout the stirring time (45 min). We label these as pre-heated suspensions. In the second protocol, the temperature of the suspension was set after stirring of the sample was stopped. These are labelled as the post-heated suspensions. 
	
	
	\subsection{2.2. Dynamic Light Scattering}
\begin{figure}[ht]
		\renewcommand{\figurename}{Figure}
		\includegraphics[width=5.5in]{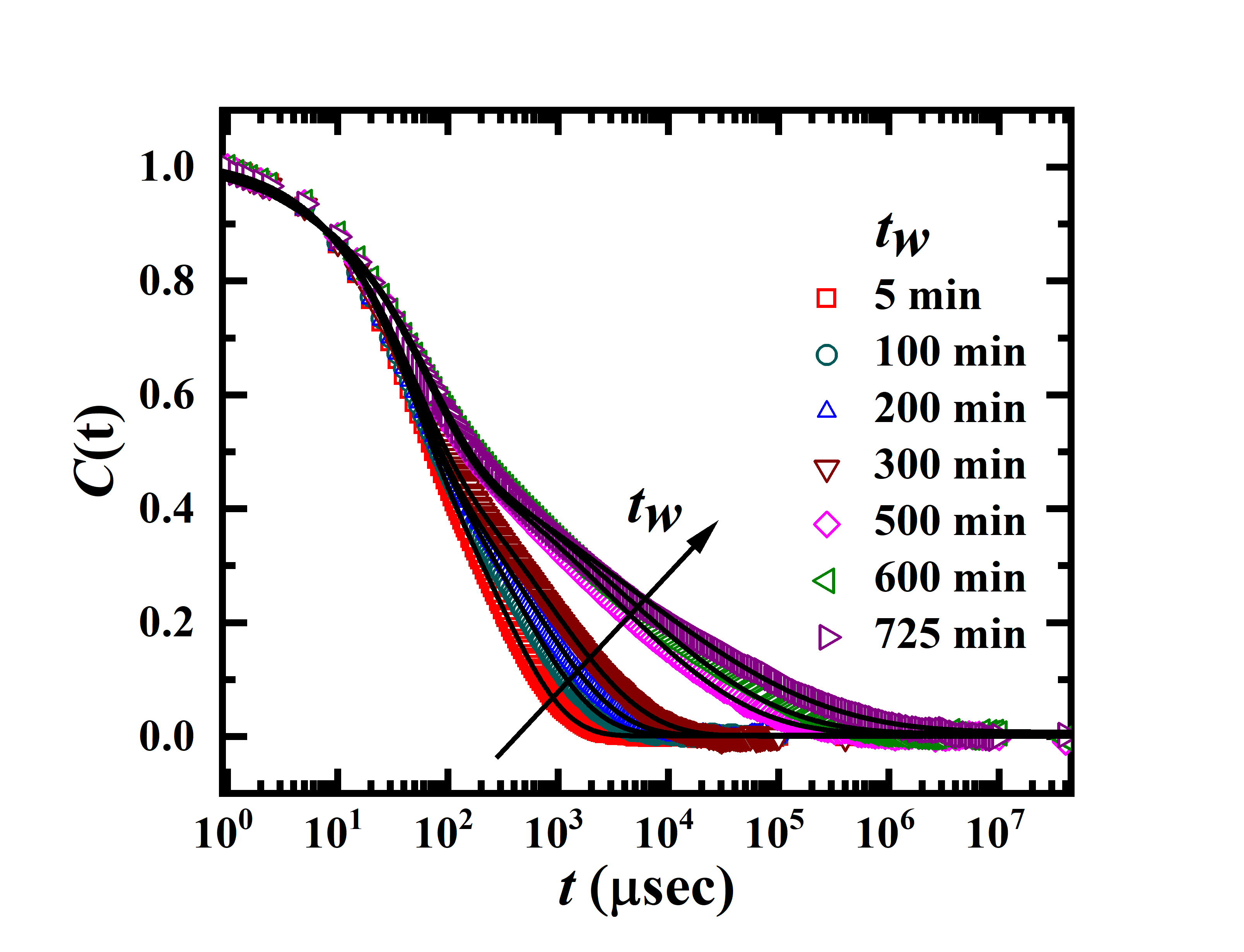}
		\caption{ Normalized autocorrelation functions, $C(t)$, {\it vs.} delay time, $t$, at $25^{\circ}$C for 12.2 mM (2.8\% w/v) aqueous Laponite suspensions plotted for different suspension ages. The solid lines are fits to eq \ref{2stepdecay}.}
		\label{Ct}
	\end{figure}

	Dynamic light scattering experiments were performed using a Brookhaven Instruments Corporation (BIC) setup whose details are given elsewhere \cite{D_Saha_2014,D_Saha_2015}. A glass cuvette filled with the sample was held in a refractive index matching bath filled with decaline. For the DLS experiments, a freshly prepared sample was filtered into the glass cuvette through a Millipore filter of diameter 0.45 $\mu$m using a syringe. The increase in the sample waiting time, $t_{w}$, was monitored continuously from the time at which filtration was stopped ($t_{w}$ = 0 at the completion of the filtration procedure). The cuvette was sealed and quickly placed in the DLS instrument to record the decays of the intensity autocorrelation functions with increasing $t_{w}$. The temperature of the sample was controlled with a temperature controller equipped with a water circulation unit. A digital autocorrelator was used to measure the intensity autocorrelation function of the scattered light: $g^{(2)}(q,t) = \frac{<I(q,0)I(q,t)>}{<I(q,0)>^{2}} =  1+ A|g^{(1)}(q,t)|^{2}$ \cite{B. J. Berne_1975}. Here $q$, $I(q,t)$, $g^{(1)}(q,t)$ and $A$ are the scattering wave vector, the intensity at a delay time $t$, the normalized electric field autocorrelation function and the coherence factor, respectively. The scattering wave vector $q$ is related to the scattering angle $\theta$, $q=(4\pi n/\lambda)\sin(\theta/2)$, where $n$ and $\lambda$ are the refractive index of the medium and the wavelength of the laser, respectively. The decays of the normalized intensity autocorrelation functions,  $C(t) = \frac{g^{(2)}(q,t)-1}{A}$, were measured for Laponite suspensions at different $t_{w}$. The normalized autocorrelation functions for 12.2 mM (2.8\% w/v) aqueous Laponite suspensions at several $t_{w}$ are plotted $vs.$ delay time $t$ in Figure \ref{Ct}.
	
	The normalized autocorrelation functions, $C(t)$ plotted in Figure \ref{Ct}, fit best to functions representing two-step decays for the entire range of waiting times explored. This indicates the presence of two relaxation time-scales corresponding to two distinct dynamical processes which can be fitted to the following equation \cite{D_Saha_2014,D_Saha_2015,P_Gadige_2017}
	\begin{equation}
		C(t) = [a\exp\{-t/\tau_{1}\} + (1-a) \exp\{-(t/\tau_{ww})^\beta\}]^2
		\label{2stepdecay}
	\end{equation}
	Fits to the above equation were used to extract $\tau_{1}$ and $\tau_{ww}$ , identified as the time-scales associated with the $\beta$- and $\alpha$- relaxation processes respectively. The faster $\beta$- relaxation involves the rattling motion of each particle inside its cage. The slower $\alpha$- relaxation corresponds to the time-scale at which the particle diffuses out of its cage in a process facilitated by the cooperative rearrangements of the surrounding particles. The parameters $a$ and $(1-a)$ represent, respectively, the relative strengths of $\beta$- and $\alpha$- relaxation processes, while $\beta$ in eq \ref{2stepdecay} is a stretching exponent that quantifies the distribution of the $\alpha$- relaxation time-scales. The average $\alpha$- relaxation time can be defined as $<\tau_{ww}> = (\frac{\tau_{ww}}{\beta})\Gamma(\frac{1}{\beta})$ \cite{C. P. Lindsey_1980}, where $\Gamma$ is the Euler Gamma function. The present work studies the approach of Laponite suspensions towards kinetic arrest by systematically monitoring the evolution of the average $\alpha$- relaxation time, <$\tau_{ww}$>, with increasing sample waiting time, $t_{w}$, in the presence of several additives and at different suspension temperatures. The diffusive nature of the dynamics even in the presence of the additives is evident from Figures S1 and S2 of Supporting Information. 

	\subsection{2.3. Rheology}
	Rheological measurements were performed using a stress controlled Anton Paar MCR $501$ rheometer. A double gap geometry with a gap of 1.886 mm, an effective length of 40 mm and requiring sample volume of $3.8$ mL was used for dilute suspensions. For dense suspensions, a cone-plate geometry having cone radius $r_{c}$ = 12.491 mm, cone angle $0.979^{\circ  }$, measuring gap $d$ = 0.048 mm and requiring sample volume of $0.07$ mL was used. For the rheological measurements, the sample was loaded immediately after preparation. Loading memory effects were erased by shear melting the sample at a high shear rate of 1500 $s^{-1}$ for 3 minutes to achieve a reproducible starting point for all experiments. The increase in the sample waiting time, $t_{w}$, was monitored continuously from the time shear melting was stopped. The temperature of the sample was controlled using a water circulation system (Viscotherm VT2). Silicone oil of viscosity 5cSt was used as a solvent trap oil to avoid water evaporation. The viscoelastic properties of aqueous  Laponite suspensions in the presence of different additives and at different suspension temperatures were then investigated by performing oscillatory amplitude sweep experiments. By varying the strain amplitude, $\gamma$, from 0.1\% to 500\%, the storage modulus, $G'$, and loss modulus, $G''$, were measured at a constant angular frequency, $\omega$, of 6 rad/s. 
	
	\subsection{2.4. Cryogenic Scanning Electron Microscopy}
	Aqueous Laponite suspensions in the presence of dissociating and non-dissociating molecules were visualized using a field-effect scanning electron microscope from Carl Zeiss with an electron beam strength of 5 kV. The samples were loaded in capillary tubes (Capillary Tube Supplies Ltd, UK) with a bore size of 1 mm using capillary flow. The ends of the capillaries were sealed. Samples were kept undisturbed before imaging and were then vitrified using liquid nitrogen slush at temperature -$190^{\circ}$C. The vitrified samples were cut and sublimated for 15 min at -$90^{\circ}$C and then coated with a thin layer of platinum (thickness approximately 1 nm) in vacuum conditions using a cryo-transfer system (PP3000T from Quorum Technologies). The cryo-chamber temperature was kept at -$190^{\circ}$C. Backscattered secondary electrons were used to produce surface images of the samples. ImageJ (Java 1.8.0\_172, developed by Wayne Rasband, NIH, US) was used to analyze the cryo-SEM images.

	\section{3. RESULTS AND DISCUSSIONS}
	
	\subsection{3.1. Microscopic aging dynamics in the presence of additives evaluated using dynamic light scattering}
	\begin{figure}[ht]
		\renewcommand{\figurename}{Figure}
		\includegraphics[width=6.4in]{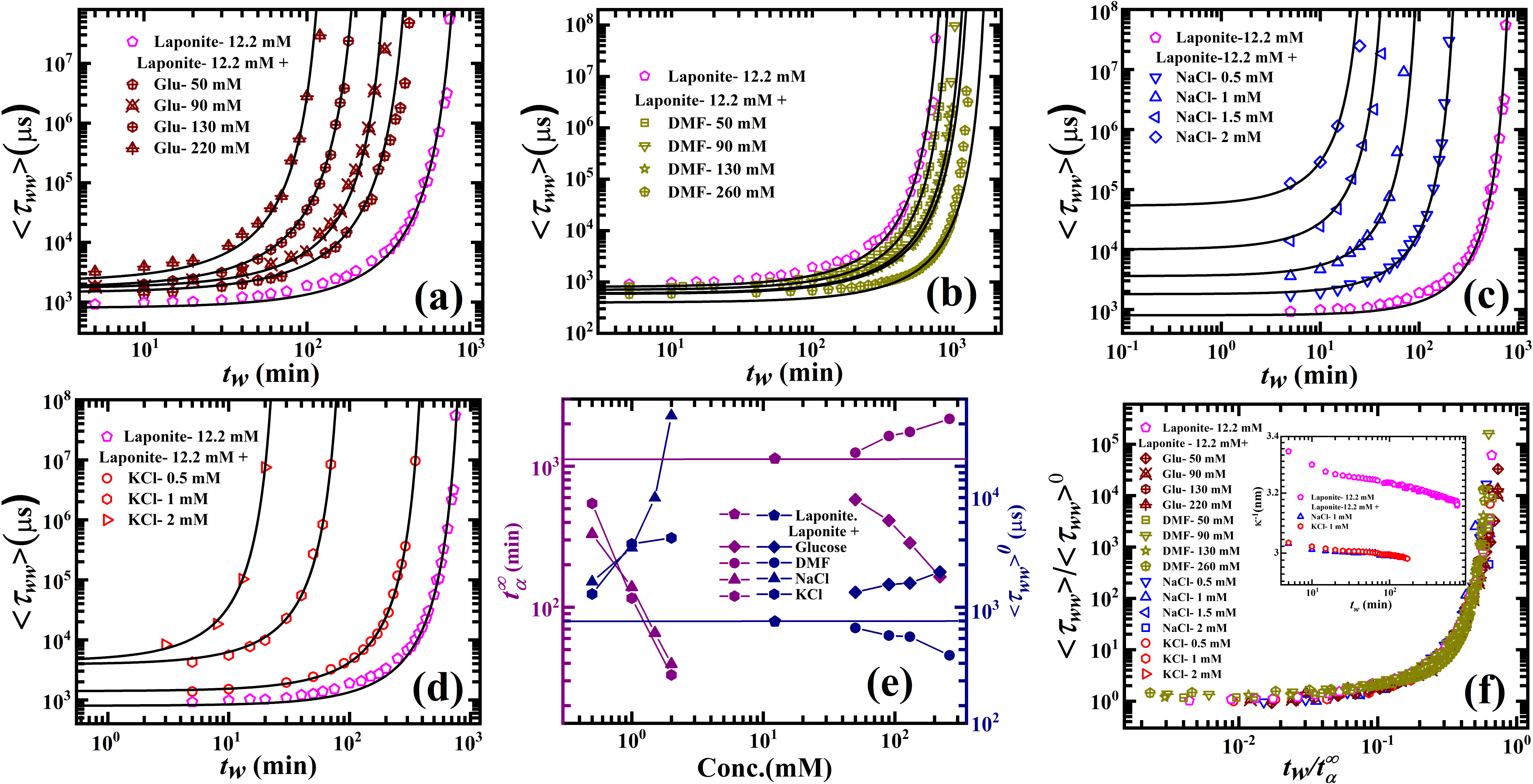}
		\caption{Mean slow relaxation time <$\tau_{ww}$> $vs$. waiting time $t_{w}$ for 12.2 mM aqueous Laponite suspensions without and with different concentrations of additives (a) glucose (Glu), (b) DMF, (c) NaCl and (d) KCl. The solid lines are fits to eq \ref{vft}. (e)  Horizontal and vertical shift factors, $t_{\alpha}^{\infty}$ (Vogel time) and $<\tau_{ww}>^{0}$, of aqueous Laponite suspensions for different concentrations of additives. (f) Superposition of normalized mean slow relaxation times <$\tau_{ww}$>/$<\tau_{ww}>^{0}$ $vs$. normalized waiting times $t_{w}/t_{\alpha}^{\infty}$ for aqueous Laponite suspensions without and with different concentrations of additives. Inset shows Debye screening lengths ($\kappa^{-1}$) of pure 12.2 mM aqueous Laponite suspension and 12.2 mM Laponite suspensions in the presence of 1 mM NaCl and KCl as a function of waiting time $t_{w}$.}	
		\label{DLS_additive}
	\end{figure}
	
	Figure \ref{DLS_additive} shows the evolution of the slow $\alpha$- relaxation time <$\tau_{ww}$> with waiting time, $t_{w}$, estimated from fits of the DLS intensity autocorrelation data to eq \ref{2stepdecay},  for 12.2 mM (2.8\% w/v) aqueous Laponite suspensions (${\bf \color{magenta}{\pentagon}}$) at 25$^\circ$C in the presence of different concentrations of dissociating and non-dissociating additive molecules in the dispersion medium. We observe that when the concentration of glucose is increased, <$\tau_{ww}$> increases rapidly as waiting time $t_{w}$ is increased (Figure \ref{DLS_additive}a). It is well known that the aging dynamics of aqueous Laponite suspensions is an osmotic pressure driven phenomenon \cite{Y_M_Joshi_2007,D_Saha_2015}. The enhancement of suspension aging in the presence of glucose, manifested by the rapid time-evolution of $<\tau_{ww}>$ (Figure \ref{DLS_additive}(a)), can be explained by considering the enhanced osmotic pressure gradients that arise due to the formation of tight hydration shells around glucose molecules \cite{M_A_Darwish_2014,V_Gekas_1998}. On the other hand, when DMF of different concentrations (shown in dark yellow symbols in Figure \ref{DLS_additive}(b)) is added to Laponite suspensions, the divergence of <$\tau_{ww}$> is delayed when compared to Laponite suspensions without any additives (${\bf \color{magenta}{\pentagon}}$) at the same temperature. The addition of DMF molecules disrupts hydrogen bonds in water, with the water molecules forming hydrogen bonds with the oxygen sites of DMF \cite{Y_Lei_2003,J_M_M_Cordeiro,P_M_Geethu_2018}. This reduces the diffusion of bulk water molecules into the intra-gallery spaces of the Laponite tactoids and results in a reduction in the osmotic pressure gradients and a retardation in the observed aging dynamics.
	
	\subsection*{3.2. Electrostatic interactions dominate in the presence of dissociating additives, while dispersion medium structure dominates in the presence of non-dissociating additives.}
	
	It is seen that when the concentrations of NaCl (shown in blue symbols in Figure \ref{DLS_additive}(c)) and KCl (shown in red symbols in Figure \ref{DLS_additive}(d)) in the suspension  are increased, <$\tau_{ww}$> increases very rapidly with waiting time $t_{w}$. The Debye screening lengths of aqueous Laponite suspensions are calculated using the expression: $\kappa^{-1}=(\epsilon_{0}\epsilon_{r}k_{B}T/\Sigma_{i}(z_{i}e)^2n_{i})^{1/2}$ \cite{Israelachvili_2010}, where $\epsilon_{0}$, $\epsilon_{r}$, $k_{B}$, $z_{i}$, $n_{i}$ and $e$ are the permittivity of free space, relative permittivity, Boltzmann constant, valency of the $i$th ion, total number density of the $i$th ion and electron charge respectively. The details of the calculation for Debye screening lengths are provided in section 1(b) of the SI. It is observed that the Debye screening lengths, $\kappa^{-1}$, of aqueous Laponite suspensions decrease continuously with waiting time which indicates the continuous dissociation of sodium ions from the Laponite faces. The addition of NaCl in the dispersion medium leads to a further decrease in $\kappa^{-1}$ (inset of Figure \ref{DLS_additive}f), indicating an increase in the induced inter-particle electrostatic attractions in the Laponite suspensions, presumably due to OC and HOC associations \cite{Samim_2016} of the Laponite platelets. The water molecules form tight hydration shells around Na$^{+}$ \cite{J_Cannon_2012} which increases osmotic pressure gradients in the suspension. An increase in both inter-particle electrostatic attractions and osmotic pressure gradients contribute to the observed acceleration in the aging dynamics of Laponite suspensions in the presence of NaCl (Figure \ref{DLS_additive}c). When KCl (widely regarded as a chaotrope and shown in red symbols) is added to the medium, the aging dynamics of the suspensions are seen to accelerate inspite of the chaotropic action of K$^+$ on the dispersion structure (Figure \ref{DLS_additive}d). The accelerated aging dynamics is attributed to a decrease in the Debye screening lengths (inset of Figure \ref{DLS_additive}f) due to the participation of K$^{+}$ ions in the EDL around each Laponite particle. This leads us to conclude that in spite of the kosmotropic nature of Na$^{+}$ and chaotropic nature of K$^{+}$ ions, electrostatics is the driving force behind the observed accelerated suspension aging in the presence of dissociating salts such as  NaCl and KCl. Clearly, electrostatic interactions dominate over any changes arising from structural modifications in suspensions containing dissociating additives.	
	
	The solid lines in Figures \ref{DLS_additive}a-d are fits to the following equation  \citealp{P_Gadige_2017,D_Saha_2014,D_Saha_2015}
	\begin{equation}
		<\tau_{ww}> = <\tau_{ww}>^{0}\exp[\frac{Dt_{w}}{t_{\alpha}^{\infty} - t_{w}}]
		\label{vft}
	\end{equation}
	where $D$ is the fragility parameter and $<\tau_{ww}>^{0}$ is the mean $\alpha$- relaxation time when $t_w$ $\rightarrow$ 0. The parameter $t_{\alpha}^{\infty}$ is identified as the Vogel time or waiting time at which the suspension achieves a kinetically arrested state \cite{D_Saha_2014}. The time-scales $t_{\alpha}^{\infty}$ and $<\tau_{ww}>^{0}$ for aqueous Laponite suspensions with and without additives are extracted by fitting the data to eq \ref{vft} and  are plotted in Figure \ref{DLS_additive}e. In Figure \ref{DLS_additive}f, we superpose the <$\tau_{ww}$> {\it vs.} $t_w$ data for all the suspensions (Figures \ref{DLS_additive}a-d) on a universal curve by dividing the horizontal and vertical axes by $t_{\alpha}^{\infty}$ and $<\tau_{ww}>^{0}$ respectively. The self-similar curvatures of the data indicate a common underlying mechanism in the aging of Laponite suspensions with and without additives. The observed increase in $<\tau_{ww}>^{0}$ (shown in dark navy symbols in Figure \ref{DLS_additive}e) and the simultaneous decrease in $t_{\alpha}^{\infty}$ (shown in dark purple symbols in Figure \ref{DLS_additive}e) in the presence of glucose, NaCl and KCl point to the increased confinement of Laponite particles in deep potential energy wells and a rapid acceleration in the aging dynamics. The addition of DMF, on the other hand, resuls in a decrease in $<\tau_{ww}>^{0}$ and an increase in $t_\alpha^\infty$, indicating shallower potential wells and retarded aging dynamics. 
	
	Figure S3 of the SI shows the conductivity of aqueous Laponite suspensions with and without additives. Owing to the uncharged nature of glucose molecules, we see a decrease in the conductivity of the suspension in the presence of glucose. The enhanced aging dynamics of Laponite suspensions in the presence of glucose is therefore understood in terms of enhanced osmotic pressure gradients arising from the formation of hydration shells. The addition of DMF has been shown by us to delay aging (Figure \ref{DLS_additive}(b)).  Although DMF is known to be conducting \cite{M_Y_Onimisi_2016}, the disruption of water structure upon the addition of DMF in Laponite suspensions clearly dominates over any  changes in interparticle electrostatics. Since glucose and DMF molecules do not participate in the EDL, these results establish the importance of structural changes in the dispersion medium in determining the suspension dynamics. EDLs of Laponite suspensions shrink in the presence of NaCl and KCl (inset of Figure \ref{DLS_additive}f). While the present results confirm earlier observations of accelerated aging in the presence of NaCl \cite{D_Saha_2015,K_Suman_2018}, we note the same effect even in the presence of KCl, a dissociating chaotropic molecule. As discussed earlier, the active participation of K$^{+}$ in the EDL results in the dominance of electrostatic screening over any disruption of medium structure that KCl may cause.
	
	\begin{figure}[ht]
		\renewcommand{\figurename}{Figure}
		\includegraphics[width=6.4 in]{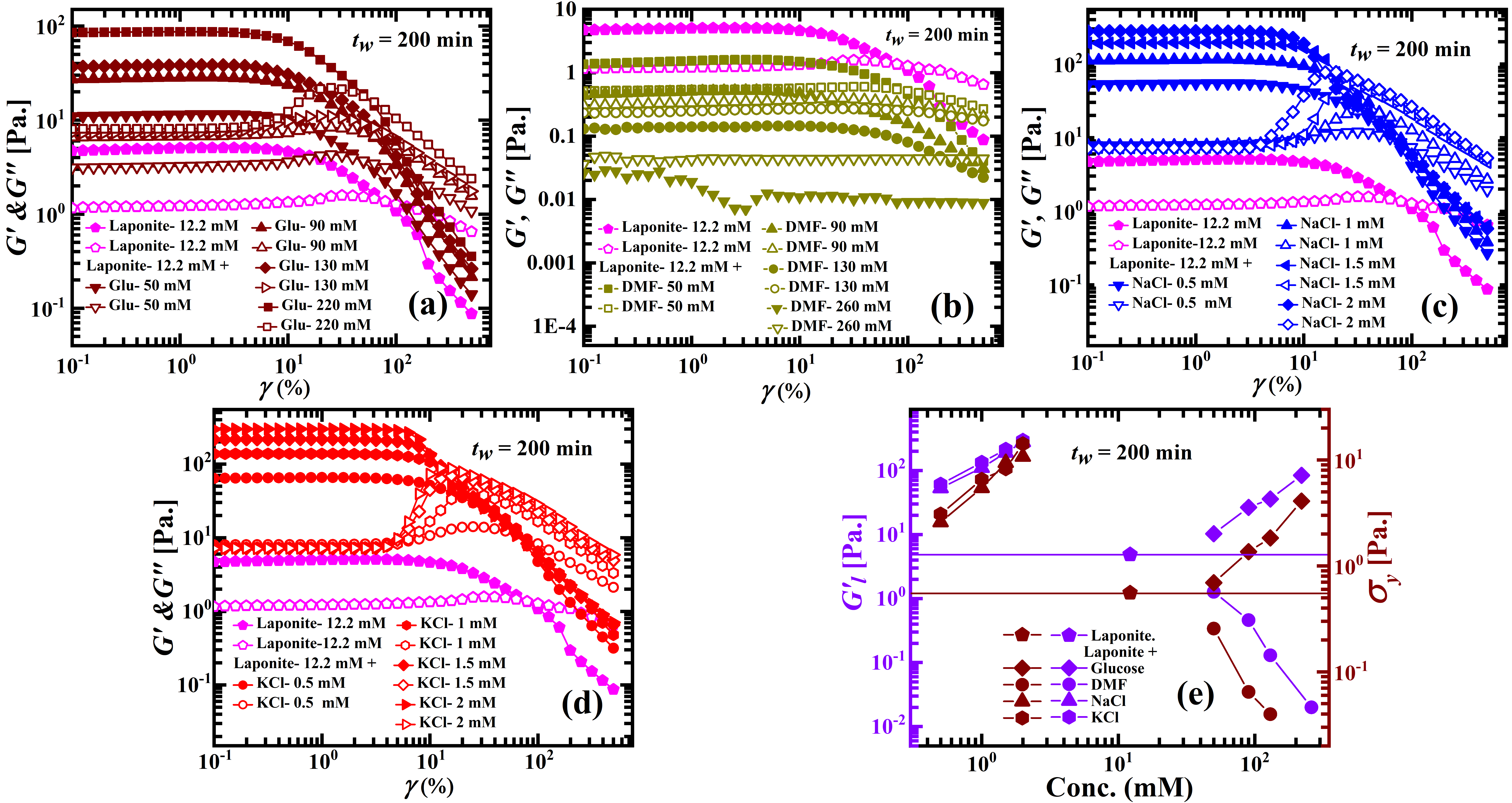}
		\caption{Storage modulus $G'$ (solid symbols) and loss modulus $G''$ (open symbols) $vs.$ applied oscillatory strain amplitude $\gamma$ for 12.2 mM aqueous Laponite suspensions in the presence of different concentrations of additives (a) glucose (Glu), (b) DMF, (c) NaCl and (d) KCl at a waiting time $t_w$ = 200 min. (e) Linear modulus and yield stresses, $G'_{l}$ and $\sigma_{y}$, of Laponite suspensions without and with additives at $t_w$ = 200 min.}	
		\label{AS_additives}
	\end{figure}
	\subsection*{3.3. Strong interparticle interactions result in structured suspensions with enhanced elasticity and yield stresses}
	We next perform oscillatory strain amplitude sweep rheology experiments to study the viscoelastic properties of aqueous Laponite suspensions with and without additives. The viscoelastic moduli ($G'$ and $G''$) of 12.2 mM aqueous Laponite suspensions ($t_w$ = 200 min) with different concentrations of additives are plotted as a function of applied strain amplitude in Figures \ref{AS_additives}a-d. At small values of the strain (LVE regime), the elastic modulus, $G'$, and the viscous modulus, $G''$, are independent of the applied strain, with $G'$ > $G''$ for all concentrations of added, glucose, NaCl and KCl and for samples with 50 and 90 mM DMF . The suspensions start yielding with an increase in the applied strain amplitude which is characterized by a monotonic decrease in $G'$, while $G''$ shows a peak before also decreasing. These are typical features of soft glassy systems with the suspensions showing predominantly fluid-like characteristics indicated by $G''$ > $G'$ at very high strain amplitudes. In contrast, for Laponite suspensions with very high DMF concentrations (130 mM and 260 mM DMF in Figure \ref{AS_additives}b), $G"$ > $G'$ in the entire strain window, signifying liquid-like behavior.
	
	In Figure \ref{AS_additives}e, we plot the linear modulus, $G'_{l}$ (the magnitudes of $G'$ extracted at very low applied strain amplitude $\gamma$, shown in violet symbols), and yield stresses, $\sigma_{y}$ (shown in wine symbols), of aqueous Laponite suspensions with and without additives at $t_{w}$ = 200 min. The yield stresses are calculated from amplitude sweep data following the method proposed by Laurati {\it et} {\it al.} \cite{Laurati_2011}. The details of the analysis and a representative plot are presented in section 1(c) and Figure S4 of the SI. Both $G'_{l}$ and $\sigma_{y}$, which estimate the strength of the underlying sample microstructures \cite{Samim_2016}, increase with increasing concentrations of glucose, NaCl and KCl, and decrease with increasing DMF concentration. The rheological results are therefore consistent with the DLS results reported earlier. In Figure S5 of the SI, we normalize the storage and loss moduli data measured in oscillatory strain sweep rheological experiments by dividing the moduli at different strain amplitudes with those measured at a strain ($\gamma$ = 0.5\%) in the LVE regime \cite{G_Y_H_Choong_2013}. As expected, the microscopic dynamics and rheological responses in the limit of low strains collapse well. In contrast, the mechanical responses of Laponite suspensions at high strains are sensitive to the presence of additives.


	\begin{figure}[ht]
		\renewcommand{\figurename}{Figure}
		\includegraphics[width=6.2in]{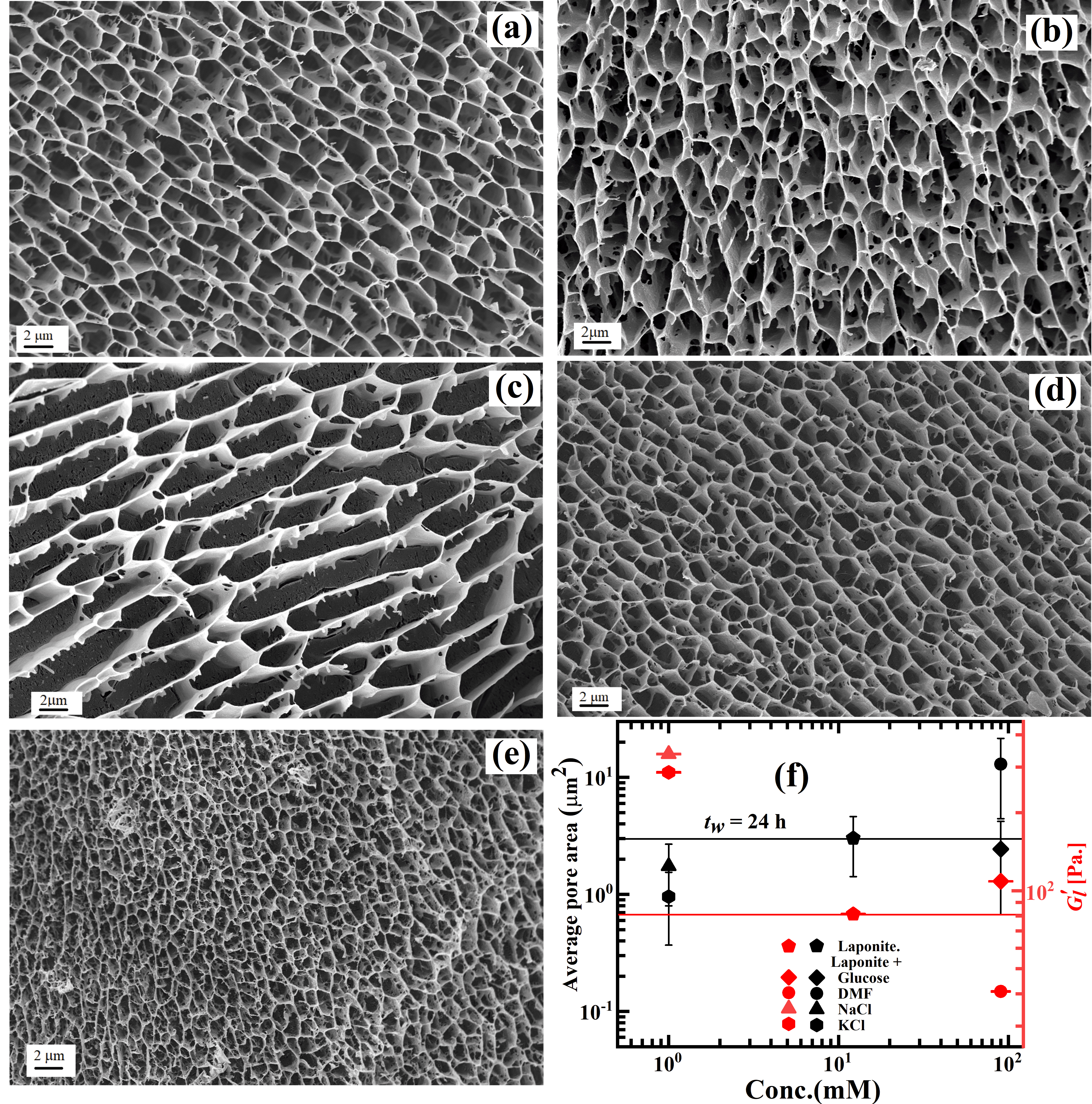}
		\caption{Representative cryo-SEM micrograph for (a) 12.2 mM aqueous Laponite suspension and Laponite suspensions in the presence of (b) 90 mM glucose, (c) 90 mM DMF, (d) 1 mM NaCl and (e) 1 mM KCl at $t_{w}$ = 24 h. (f) Average pore area (black) and linear modulus (red) for aqueous Laponite suspensions without and with additives at $t_{w}$ = 24 h.}	
		\label{Cryo-SEM}
	\end{figure}
	\subsection*{3.4. Decrease in average pore area of the self-asssembled suspension structures increases suspension elasticity}	
	We perform cryo-SEM experiments to investigate the morphologies of the samples studied here. Figures \ref{Cryo-SEM}a-e display the microstructures of aqueous Laponite suspensions without and with additives at a waiting time of 24 h. Honeycomb like network structures are seen in all samples, with pore sizes that are sensitive to the additive present in the suspension. We note the existence of holes on the flat surfaces of the network branches, indicating the possibility of overlapping coin configurations (OC) \cite{B. Jonsson_2008} of the Laponite platelets \cite{Samim_2016}. The magnified cryo-SEM micrographs of the sample microstructures are provided in Figure S6.  
	
	We adopt a protocol used earlier \cite{Samim_2016} to quantify the porous microstructures by estimating pore areas (data at $t_{w}$ = 24 h is shown in black symbols in Figure \ref{Cryo-SEM}f) and branch thicknesses (Figure S7). In the analysis of cryo-SEM images, the presence of vitrified water on the structures can lead to the overestimation of the network branch thickness and a simultaneous underestimation of average pore area. However, since the sublimation time (15 min) after cutting the vitrified samples is identical in all the experiments, an equal sublimation-depth is expected for all the samples studied using cryo-SEM. We correlate network morphologies with their mechanical responses obtained in rheological experiments. The linear modulus, $G'_{l}$ (magnitude of $G'$ of all the samples at $t_{w}$ = 24 h extracted at very low applied strain amplitude $\gamma$ (Figure S8)), are plotted in Figure \ref{Cryo-SEM}f (shown in red symbols). While the branch thicknesses of the samples remain unchanged in all the samples studied here, the average pore area decreases for Laponite suspensions in the presence of glucose ($\Diamondblack$), NaCl ($\blacktriangle$) and KCl ($\varhexagonblack$). Figure \ref{Cryo-SEM}f displays a simultaneous increase in the elasticity of these samples. In contrast, an increase in average pore area and a simultaneous decrease in elasticity is noted when DMF of concentration 90 mM ($\CIRCLE$) is added to Laponite suspensions (Figure \ref{Cryo-SEM}f). Many small dangling branches are observed in the walls of the sample microstructures in Figure \ref{Cryo-SEM}c, which indicates incomplete structure formation and retarded aging dynamics in the suspension. In Figure S9, we show the microstructures of aqueous Laponite suspensions in the presence of all the additives at $t_{w}$ = 200 min. When glucose, NaCl and KCl are added to Laponite suspensions, network structures, with pore areas larger than those observed in the same samples at a longer waiting time ($t_{w}$ = 24 h), are seen (Figure S9f). This indicates the continual aging of the suspension structures. Clearly, aqueous Laponite suspensions age faster in the presence of dissociating molecules such as NaCl and KCl due to the dominance of inter-particle electrostatic interactions over alterations in the dispersion medium structure. The contribution of microstructure dominates only in the case of uncharged molecules, with the kosmotropic glucose and the chaotropic DMF respectively causing acceleration and delay in suspension aging. 
	
	\subsection{3.5. In temperature-controlled experiments that modify hydrogen bonding in the dispersion medium, electrostatics determines suspension aging}
	\begin{figure}[ht]
		\renewcommand{\figurename}{Figure}
		\includegraphics[width=6.4in]{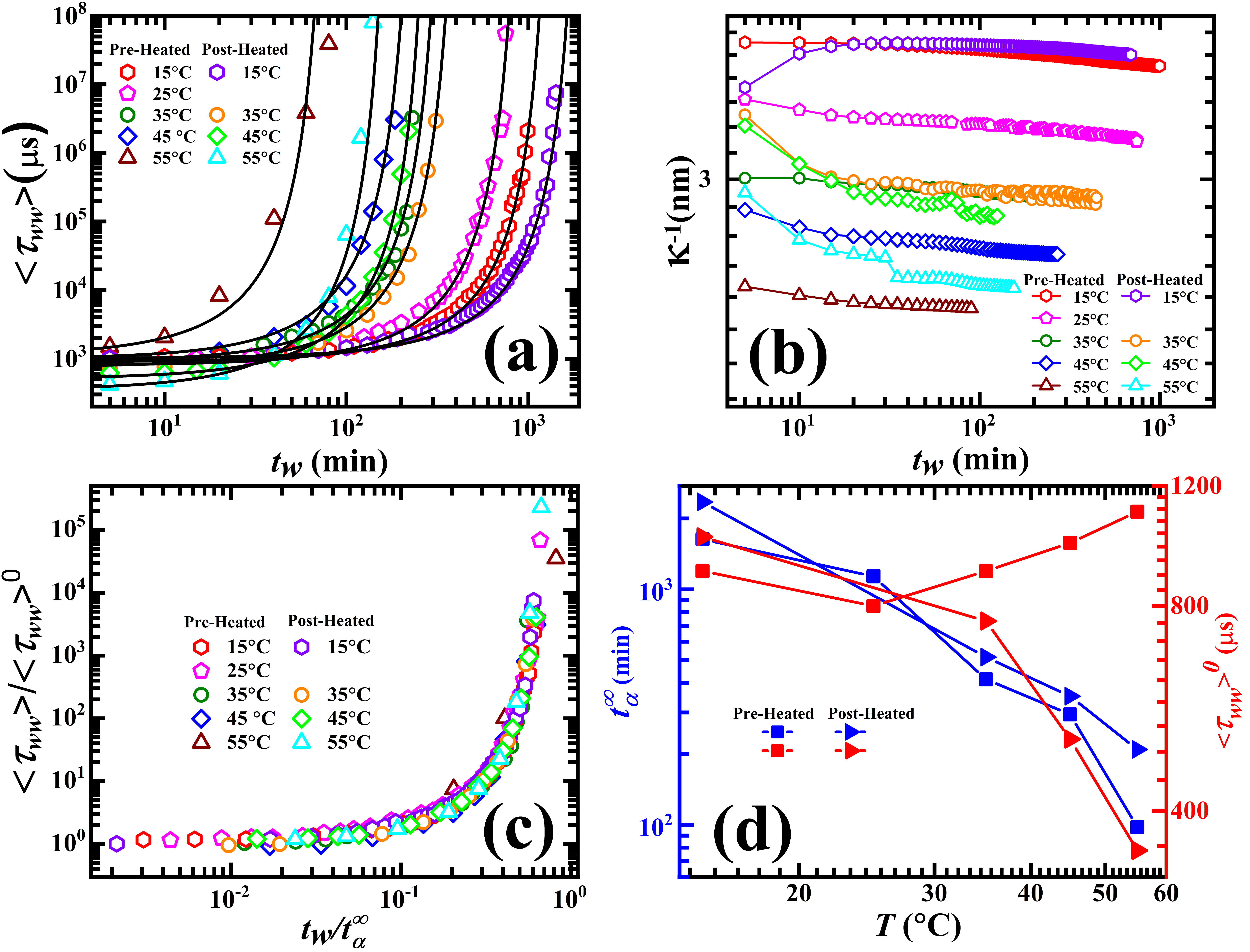}
		\caption{(a) Mean slow relaxation time <$\tau_{ww}$> {\it vs.} waiting time $t_{w}$ for pre-heated and post-heated aqueous Laponite suspensions of concentration 12.2 mM. The solid lines are fits to eq \ref{vft}. (b) Debye screening lengths ($\kappa^{-1}$) of pre- and post-heated suspensions as a function of $t_{w}$. (c) Superposition of normalized mean slow relaxation times <$\tau_{ww}$>/$<\tau_{ww}>^{0}$ $vs.$ normalized waiting times $t_{w}/t_{\alpha}^{\infty}$. (d) Horizontal and vertical shift factors, $t_{\alpha}^{\infty}$ (Vogel time) and $<\tau_{ww}>^{0}$, of pre- and post-heated suspensions.}	
		\label{DLS_Temp}
	\end{figure}

	Figure \ref{DLS_Temp}a plots the dependence of the mean slow $\alpha$- relaxation time, <$\tau_{ww}$>, estimated from fits of intensity autocorrelation functions (eq 1) acquired in  DLS experiments, on $t_{w}$, for pre- and post-heated Laponite suspensions. We observe that higher suspension temperatures result in accelerated aging dynamics. This kosmotrope-like effect of temperature on the aging dynamics in a post-heated suspension has been reported in an earlier work \cite{D_Saha_2015}. We note that while increasing temperature disrupts hydrogen bonds, there is a simultaneous enhancement in the inter-particle electrostatic attraction, arising due to decrease in the Debye screening length $\kappa^{-1}$ ( Figure \ref{DLS_Temp}b, calculated from conductivity measurements plotted in Figure S10 of the SI). At increased temperatures, therefore, electrostatics dominates over disruption in medium structure while determining suspension dynamics. Interestingly, we observe a comparatively more rapid enhancement in the aging dynamics of aqueous Laponite suspensions when temperature of water is raised before the addition of Laponite particles (Figure \ref{DLS_Temp}a). The enhanced aging in pre-heated samples when compared to post-heated ones arises from the larger decrease in Debye lengths in the former case (Figure \ref{DLS_Temp}(b)) due to an increased dissociation of Na$^{+}$ in the former experimental protocol. The superposition plot of the <$\tau_{ww}$> data (Figure \ref{DLS_Temp}c), shows self-similar behavior, thereby revealing an indistinguishable aging mechanism of aqueous Laponite suspensions at all temperatures \cite{D_Saha_2015}. The horizontal and vertical shift factors, $t_{\alpha}^{\infty}$ and $<\tau_{ww}>^{0}$ (extracted by fitting the data in Figure \ref{DLS_Temp}a to eq \ref{vft}), are plotted in Figure \ref{DLS_Temp}d. The observed decrease in $t_{\alpha}^{\infty}$ (displayed in Figure \ref{DLS_Temp}d using blue symbols) with increase in suspension temperature indicates accelerated aging dynamics due to accelerated structural buildup in the suspensions. Lower values of $t_{\alpha}^{\infty}$ (${\bf \color{blue}{\blacksquare}}$) in pre-heated suspensions when compared to post-heated ones (${\bf \color{blue}{\blacktriangleright}}$) quantitatively verify the presence of accelerated aging dynamics in the former case. While a monotonic decrease in $<\tau_{ww}>^{0}$ with temperature in post-heated suspensions (${\bf \color{red}{\blacktriangleright}}$) indicates the chaotropic effect of temperature at time $t_w$ $\rightarrow$ 0, we observe a non-monotonic behavior in $<\tau_{ww}>^{0}$ in pre-heated suspensions (${\bf \color{red}{\blacksquare}}$). A higher value of $<\tau_{ww}>^{0}$ at 15$^\circ$C in the pre-heated suspension reveals the dominance of an initial kosmotropic behavior. In contrast to post-heated suspensions, $<\tau_{ww}>^{0}$ of the pre-heated samples increase with temperature at {\it T }> 25${^\circ}$C due to rapid acceleration in the aging dynamics under these conditions. The influence of water structure on suspension dynamics at larger $t_w$ is minimal in most of these experiments.

	\begin{figure}[h]	
		\renewcommand{\figurename}{Figure}
		\includegraphics[width=6.5in]{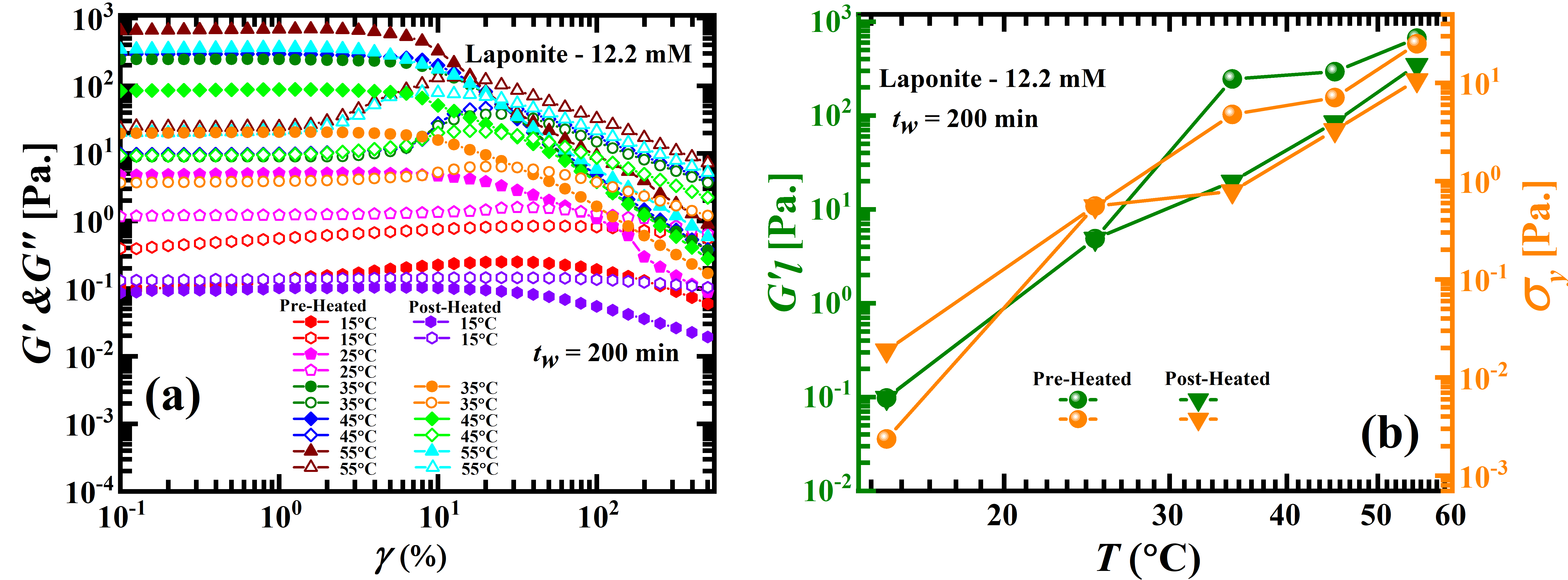}
		\caption{(a) Storage modulus $G'$ (solid symbol) and loss modulus $G''$ (open symbol) {\it vs.} applied oscillatory strain amplitude $\gamma$ for pre- and post-heated Laponite suspensions at $t_{w}$ = 200 min. (b) Linear modulus and yield stresses, $G'_{l}$ and $\sigma_{y}$, of Laponite at $t_{w}$ = 200 min for temperatures in the range 15-55$^\circ$C both for pre- and post-heated suspensions.}	
		\label{AS_temp}
	\end{figure}
	
	Next, oscillatory amplitude sweep rheological experiments are performed to study the viscoelastic behavior of aqueous Laponite suspensions at different temperatures. Figure \ref{AS_temp}a shows the temperature dependence of the viscoelastic moduli ($G'$ and $G''$) of 12.2 mM aqueous Laponite suspensions at $t_{w}$ = 200 min as a function of applied strain amplitude, $\gamma$. The linear modulus, $G'_{l}$ (olive symbols), and yield stresses, $\sigma_{y}$ (orange symbols), of aqueous Laponite suspensions increase with temperature (Figure \ref{AS_temp}(b)) which indicate enhanced rates of structure buildup \cite{Samim_2016} and accelerated aging dynamics. The higher $G'_{l}$ and $\sigma_{y}$ values at higher temperatures in the pre-heated suspensions when compared to post-heated ones confirm faster aging in the former case and are in agreement with the results of the DLS experiments reported in Figure \ref{DLS_Temp}. Similar to observations in Laponite suspensions with externally added  additives, we note that while the microscopic dynamics and linear rheological responses collapse well at all temperatures (except at 15$^\circ$C for the pre-heated samples (Figure 5(d)) in which the kosmotropic effect dominates), the nonlinear mechanical responses are sensitive to temperature histories over the experimental strain windows (Figure S11 of the SI). We conclude that even though increase in temperature disrupts hydrogen bonds in the dispersion medium, the enhancement in inter-particle electrostatics due to increased dissolution of Na$^{+}$ from the Laponite particles dominates suspension dynamics. While the kosmotrope like effect of temperature has been demonstrated before \cite{D_Saha_2015,K_Suman_2018}, the differences between pre-heating and post-heading have never been reported in the literature to the best of our knowledge. 
	
	\section{4. CONCLUSION}      
	This study identifies the dispersion medium structure as one of the key control parameters that determines the physicochemical properties of aqueous colloidal suspensions of charged nano-clay particles. An enhancement in hydrogen bond population in the medium by adding uncharged kosmotropic molecule accelerates the suspension dynamics, increases the elastic modulus of the sample and reduces the network pore sizes of the suspension structures. In contrast, addition of uncharged chaotropic molecules which disrupt the medium structure delays the aging dynamics, reduces the elastic modulus and increases the network pore sizes. Inter-particle electrostatic interactions determine the suspension properties in the presence of charged dissociating molecules, suppressing the action of local structure of the dispersion medium. Our inferences on the particle dynamics are in agreement with the rheological properties of the samples and correlate well with the microstructural details of the suspensions visualized using cryogenic scanning electron microscopy. 
	\paragraph{}
	Rios $et$ $al$. have reported that kosmotropes promote the aggregation of hydrophobic solutes while chaotropes destabilizes them and that the same observation holds good for amorphous aggregates \cite{rios,kentaro}. It was reported that kosmotrope molecules enhance the fusion temperature of poly(vinyl alcohol) cryogels and chaotrope molecule reduce their rigidity \cite{Lozinsky}. Zoochi $et$ $al$ identified that kosmotropic substances stiffen the protein-hydration layer system while chatropes soften this layer \cite{N_C_Morales}. In our study, we also observe enhanced rigidity of Laponite suspensions when an uncharged kosmotropic molecule (glucose) is added to the dispersion medium, while we report a reduction when an uncharged chaotrope molecule (N,N-dimethylformamide) is added. Charged, dissociating molecules induce strong electrostatic interactions between clay particles and hinder the influence of local structure of the dispersion medium in determining sample properties. Despite the crucial roles of soil water content and salinity in agriculture and large scale geophysical phenomena such as river delta formation and land slides, a laboratory scale experimental investigation on the influence of water structure on the physicochemical properties of aqueous clay colloids has never been considered before. Our results, which are in agreement with previous studies of water structure on the aggregation properties of biological and physical polymer systems \cite{rios,N_C_Morales}, can be generalized to any colloidal system where the sample properties are determined by a combination of local osmotic pressure gradients and electrostatic interactions.
	\section*{Author contributions}
	CM, VTR and RB designed the experiments, CM and VTR performed the experiments and analyzed the data. All authors contributed to drafting and editing the manuscript.
	\section{ASSOCIATED CONTENT}
	\subsection{Supporting Information}
	Section 1(a) : Diffusive behavior of Laponite suspensions in the presence of different additives are shown in Figures S1 and S2. Section 1(b) : Discussion of Debye screening length calculations. Figure S3 : Plots of conductivity of Laponite suspensions in the presence of different additives. Section 1(C): Details of yield stress calculation, with representative data plotted in Figure S4. Figure S5: Plots of normalized storage and loss moduli {\it vs.} applied strain. Section 1(d), Figures S6-S8: Magnified cryo-SEM images, calculation of network branch thicknesses of Laponite suspensions and strain amplitude sweep rheological data in the presence of different additives at waiting time 24h. Figure S9: Magnified cryo-SEM images in the presence of different additives at waiting time 200 min. Section 2, Figures S10-S11: Conductivity measurements {\it vs.} sample age and normalized storage and loss moduli data {\it vs.} applied strain at different suspension temperatures.   
	
	\section{Acknowledgments}
	We thank Ananya Saha for her help during the initial stages of the experiments and K. M. Yatheendran for his help with cryo-SEM imaging. We thank Raman Research Instistute for funding our research and DST SERB (grant number EMR/2016/006757) for partial financial support.

\end{document}